\documentclass[]{jfm}
\usepackage{dcolumn}
\usepackage{bm}

\usepackage[utf8]{inputenc}
\usepackage[T1]{fontenc}
\usepackage{mathptmx}
\usepackage{etoolbox}
\usepackage{color}
\usepackage{ulem}
\usepackage{overpic}

\usepackage{subcaption} 
\usepackage{graphicx}
\usepackage{overpic}%
\usepackage{makecell}
\usepackage{multirow}
\usepackage{rotating}

\usepackage{newtxtext}
\usepackage{newtxmath}
\usepackage{natbib}
\usepackage{hyperref}
\usepackage{multirow}
\hypersetup{
    colorlinks = true,
    urlcolor   = blue,
    citecolor  = blue, 
}

\newcommand{\RomanNumeralCaps}[1]
\linenumbers

\title{Optimizing thermal convection by phase-locking circulation to wall oscillations}

\author{YaLin Zhu\aff{1},
   Jian-Chao He\aff{2}  
  \and Xi Chen\aff{1}
  }

\affiliation{
\aff{1}Institute of Fluid Mechanics, Beijing University of Aeronautics and Astronautics, 100191 Beijing, China
\aff{2}High Performance Computing Department, National Supercomputing Center in Shenzhen, 518028, Shenzhen, China

}

\begin{document}
\maketitle

\begin{abstract}

This study numerically investigates heat transfer and flow reorganization in two-dimensional Rayleigh-Bénard convection subjected to horizontal oscillation of the bottom plate, with a fixed Prandtl number $Pr=4.3$, Rayleigh numbers $Ra\in[5\times10^6,1\times10^8]$, and oscillation frequencies $f\in[0.0001,0.5]$. The imposed oscillation breaks the up--down symmetry of the classical RBC system and induces a strong frequency-dependent response in global heat transport, with the maximum Nusselt number enhancement exceeding $60\%$ compared to the uncontrolled case. Central to the control efficiency is a phase-locking mechanism: at the optimal frequency $f=f_{\text{opt}}$, the intrinsic response time of large-scale circulation (LSC), quantified by the sign-recovery of the volume-averaged angular momentum $\Omega$, locks precisely to the wall oscillation period, enabling perfectly synchronized LSC reversals. Deviations from this condition lead to a marked mismatch -- the LSC response time becomes substantially longer when $f>f_{\text{opt}}$ and significantly shorter when $f<f_{\text{opt}}$, relative to the oscillation period. In contrast, velocities within the bottom and sidewall boundary layers exhibit periodic variations always following the wall oscillations, therefore unable to distinguish the control efficiency when $f$ changes. Fourier mode analysis further shows that at optimal frequency, the single-roll mode ($M^{1,1}$) remains dominant throughout the cycle, facilitating efficient plume transport and maximizing heat transfer. At higher frequencies, the LSC cannot follow the rapid forcing, resulting in incomplete reversals; at lower frequencies, a double-roll structure ($M^{1,2}$) emerges, which reduces heat-transfer efficiency despite intensified plume emission. This frequency-locking mechanism is shown to persist for optimal controls across the investigated range of Rayleigh numbers here, thus offering insight for active control strategies in thermally driven turbulent flows.


\end{abstract}



\section{Introduction}
\label{sec:intro}

Thermal convection is ubiquitous in natural and engineering systems, driven by buoyancy forces arising from temperature gradients. Among the canonical models for buoyancy-driven turbulence, Rayleigh--B\'{e}nard convection (RBC), characterized by heating from below and cooling from above, provides a well-defined framework for studying fundamental flow phenomena such as large-scale circulation (LSC) \citep{benzi2005flow,brown2007large,brown2008model,assaf2011rare,ni2015reversals}, thermal plume dynamics, and boundary-layer evolution \citep{ahlers2009heat,lohse2010small,chilla2012new,sun2014experimental,xia2013current,xia2023tuning}.
Extensive research has established scaling relations linking dimensionless heat transfer efficiency (Nusselt number, $Nu$) to driving strength (Rayleigh number, $Ra$) and fluid properties (Prandtl number, $Pr$) and revealed distinct transitions in different flow regimes when the control parameters vary \citep{grossmann2000scaling,ahlers2009heat,iyer2020classical, Doering2020PNAS, Lohse2024ultimate, shishkina2024ultimate, tiwari2025absence}.  

Beyond the intrinsic scaling behavior of RBC, increasing attention has been devoted to flow control strategies that deliberately manipulate flow organization and boundary-layer dynamics to regulate convective heat transport. \citet{jiren2025comprehensive} classified these strategies into passive and active approaches according to whether external energy or time-dependent forcing is introduced. Passive control relies on static modifications of system geometry, boundary conditions, or orientation, exploiting the sensitivity of turbulent convection to boundary-layer stability and LSC without additional energy input. By contrast, active control employs external forcing or feedback mechanisms, such as thermal or mechanical modulation and data-driven control, to dynamically steer flow states and achieve stronger heat-transfer enhancement, albeit at the cost of additional energy expenditure and system complexity. Since the thermal boundary layer constitutes the primary bottleneck for heat transfer \citep{iyer2020classical,guo2022turbulent}, both passive and active are mostly designed to modify the instability and dynamical properties of thermal boundary layer.

Passive methods modify convective heat transport through system geometry, boundary conditions, or gravitational orientation. Geometric modifications, such as the variation of the aspect ratio \citep{gelfgat1999different} and the surface roughness \citep{wagner2015heat, zhu2017roughness, Cheng2024JFM}, reorganize flow structures and disrupt the thermal boundary layer, enhancing thermal convection. Optimized boundary temperature distributions can further tailor convective cell dynamics to improve heat transfer \citep{jiren2024optimizing,jiren2025comprehensive}. The thermal conditioning of the side wall regulates the stability of the LSC and the heat transport. Isothermal or controlled sidewall zones suppress flow reversals and stabilize the LSC, leading to more coherent thermal transport \citep{zhang2020controlling,chen2019emergence}. 
The tilt of the system modifies the buoyancy component along heated surfaces, which can raise the value of $Nu$ \citep{grunkleton2006numerical,gawas2022natural}. In rotating systems, tilt couples with Coriolis forces to reorganize the flow into more pathways efficiently \citep{novi2019rapidly}, providing a simple means to tune convective intensity without external energy. The insertion of partitions or bulk structures to manipulate flow geometry has been shown to significantly enhance heat transport through coherent plume organization, as demonstrated by \citet{bao2015enhanced} and \cite{zhang2023achieving}. Collectively, these passive approaches can enhance heat transfer by reorganizing flow structures and stabilizing circulations.

Beyond passive methods, a variety of active control strategies have been proposed for RBC. 
On the thermal forcing side, periodic modulation of bottom-plate temperature has been shown to thin the thermal boundary layer, increasing $Nu$ by up to 25\% at an optimal forcing frequency \citep{yang2020periodically}.
Steady sidewall heating has also demonstrated significant heat transfer enhancement (up to 66\%) by intensifying the LSC \citep{mac2024side}. Similarly, isothermal sidewall control can enhance heat transfer \citep{zhang2020controlling,zhang2021stabilizing},
where a two-point control configuration breaks the up-down symmetry and strengthens the LSC, increasing $Nu$ by up to 4.3\%.
In addition, mechanical forcing provides an alternative route: \citet{wang2020vibration} demonstrated that horizontal oscillation can destabilize boundary layers, resulting in heat transfer enhancements of up to 600\%, while vertical oscillation suppresses plume emission and significantly reduces heat flux \citep{wu2022vibration}. More recently, \citet{yuan2023boundary} introduced a standing-wave-type boundary deformation that disrupts boundary-layer coherence when the deformation amplitude is comparable to or exceeds the boundary-layer thickness. Under such conditions, heat transfer increases by more than 100\%, and the $Nu$ scaling approaches the ultimate regime ($Nu \sim Ra^{1/2}$). \citet{zhao2025coupled} showed that horizontal vibration combined with a rough wall disrupts recirculation zones within roughness gaps and thins the thermal boundary layer, leading to a $Nu$ increase of up to 206\%—a gain significantly exceeding the sum of individual contributions from roughness or vibration alone.
In parallel, \citet{xue2025controlling} leveraged the thermoelectric effect to actively modulate heat transfer in vertical convection by applying a non‑uniform magnetic field. The resulting Lorentz force enhances or suppresses the LSC, achieving a controllable $Nu$ enhancement of up to 60\%.

A long studied control for wall turbulence is through plate oscillation, which can attenuate near-wall streaks and suppress vortex activity by periodically perturbing the turbulence regeneration cycle \citep{quadrio2011drag, marusic2021energy, ricco2021review, Gatti2025, Yao2019reynolds, zhang2025reynolds}. This idea has also been extended to RBC with the expectation of altering the thermal boundary layer. For example, \cite{liu2025prandtl} systematically investigated the effect of Prandtl number on RBC modulated by an oscillatory bottom plate, revealing that the critical velocity for boundary layer instability scales as $\bar{V}_c \sim Pr^{0.5}$ and that heat transfer reduction can occur even when boundary layer instabilities are triggered for $Pr \leq 1$. Similarly, \cite{Yang2024PoF} examined the role of a rotating bottom endwall in a cylindrical RBC system, demonstrating that the induced meridional circulation significantly enhances heat transport in the rotation-dominated regime, with unified scaling relations $Nu \sim Ra^{0.3}\omega^{0.64}$ and $Re \sim Ra^{0.5}\omega$. More recently, \cite{ZhangLe2025JFM} set a spatiotemporal modulation for wall temperature and found different plume dynamics and heat-transfer scaling. In this study, the wall oscillation is imposed on the bottom plate, while the wall temperature is fixed as constant. This approach breaks the up--down symmetry, thus focusing on the asymmetric response of RBC under active excitations, which has not been explored before.

Here, we conducted a direct numerical simulation (DNS) to investigate how the periodic oscillation (horizontal) of the bottom plate changes heat transfer and flow organization in two-dimensional (2D) RBC. Note that although three-dimensional (3D) simulations capture the full complexity of turbulent thermal convection, 2D studies remain indispensable for identifying primary mechanisms and allowing systematic parametric exploration, due to the lower computational cost. Previous work has demonstrated that 2D RBC reproduces the key scaling behaviors of $Nu$ and $Re$ as in 3D systems \citep{van2013comparison, ZhangY2017JFM}. Moreover, the introduction of vibration in 2D thermal convection has been shown to significantly modify flow organization \citep{guo2022turbulent, Guo2024POF}. Therefore, 2D simulations provide an efficient way to investigate oscillation-driven convection. Despite these studies, less attention has been paid to the asymmetric response of RBC under active mechanical forcing, especially the possible role of phase synchronization between large-scale circulation reversals and wall oscillations. Our current work attempts to address this aspect by identifying a phase-locking mechanism that appears to govern optimal heat transfer enhancement, which may offer a different dynamical perspective for active control strategies in thermally driven turbulence.

The remainder of the paper is organized as follows. Section~2 introduces the governing equations, numerical methods, and simulation parameters. Section~3 presents the frequency-dependent heat-transfer enhancement at \(Ra = 1 \times 10^{7}\) (\S3.1), followed by an analysis of boundary-layer behavior and LSC reversal (\S3.2), and Fourier mode decomposition to elucidate the underlying flow structures (\S3.3). Section~4 extends the analysis to other Rayleigh numbers and verifies the locking mechanism for flow pattern classification. Finally, Section~5 summarizes the main findings and discusses their implications for active control of thermal convection.

\section{Equations and numerical settings}
\label{sec:equations}
The governing equations for the RBC system are derived under the  Oberbeck-Boussinesq (OB) approximation, where the density-temperature follows the linear relation $\rho(T) = \rho(T_0) \left [1-\alpha (T-T_0) \right]$ with $T_0$ the reference temperature. All fluid transport coefficients, including the thermal expansion coefficient $\alpha$, kinematic viscosity $\nu$, and thermal diffusivity $\kappa$, remain constant. This approximation couples the temperature and velocity fields through the buoyancy effect exclusively. Note that non-OB effects are excluded from our analysis, as they lead to fundamentally different thermal and flow structures in RBC systems \citep{bodenschatz2000recent, sugiyama2007non,pan2023non}.

Using characteristic scales of height $H$, temperature difference $\triangle T$, and velocity $U = \sqrt{\alpha g H \triangle T}$,
 and free-fall time $t_f = \sqrt{H/\alpha g \triangle T}$, the dimensionless RBC equations can be written as:
\begin{equation}
\frac{\partial u_i}{\partial x_i} = 0,  \label{eq:div}
\end{equation}
\begin{equation}
\frac{\partial u_i}{\partial t} + u_j \frac{\partial u_i}{\partial x_j}
    = -\frac{\partial p}{\partial x_i}
       + \frac{1}{\sqrt{Ra/Pr}}\frac{\partial^2 u_i}{\partial x_j \partial x_j}
       + \theta \delta_{i2},  \label{eq:momentum}
\end{equation}
\begin{equation}
\frac{\partial \theta}{\partial t} + u_j \frac{\partial \theta}{\partial x_j}
    = \frac{1}{\sqrt{Ra \cdot Pr}}\frac{\partial^2 \theta}{\partial x_j \partial x_j}. \label{eq:thermal}
\end{equation}
where $u_i(x_j,t)$, $\theta(x_j,t)$, and $p(x_j,t)$ are the dimensionless velocity, temperature, and pressure fields, respectively; $i,j=1,2$ indicate velocity components and spatial coordinates.
The Kronecker delta $\delta_{i2}$ activates the buoyancy term only in the vertical direction ($i=2$). The Rayleigh number is defined as $ Ra = \alpha g \Delta T H^3 / (\nu \kappa) $, where $g$ is the gravitational acceleration. The Prandtl number is defined as $ Pr = \nu / \kappa $, which represents the ratio of momentum diffusivity to thermal diffusivity.

\begin{figure}
    \centering
    \begin{overpic}[width=0.5\linewidth]{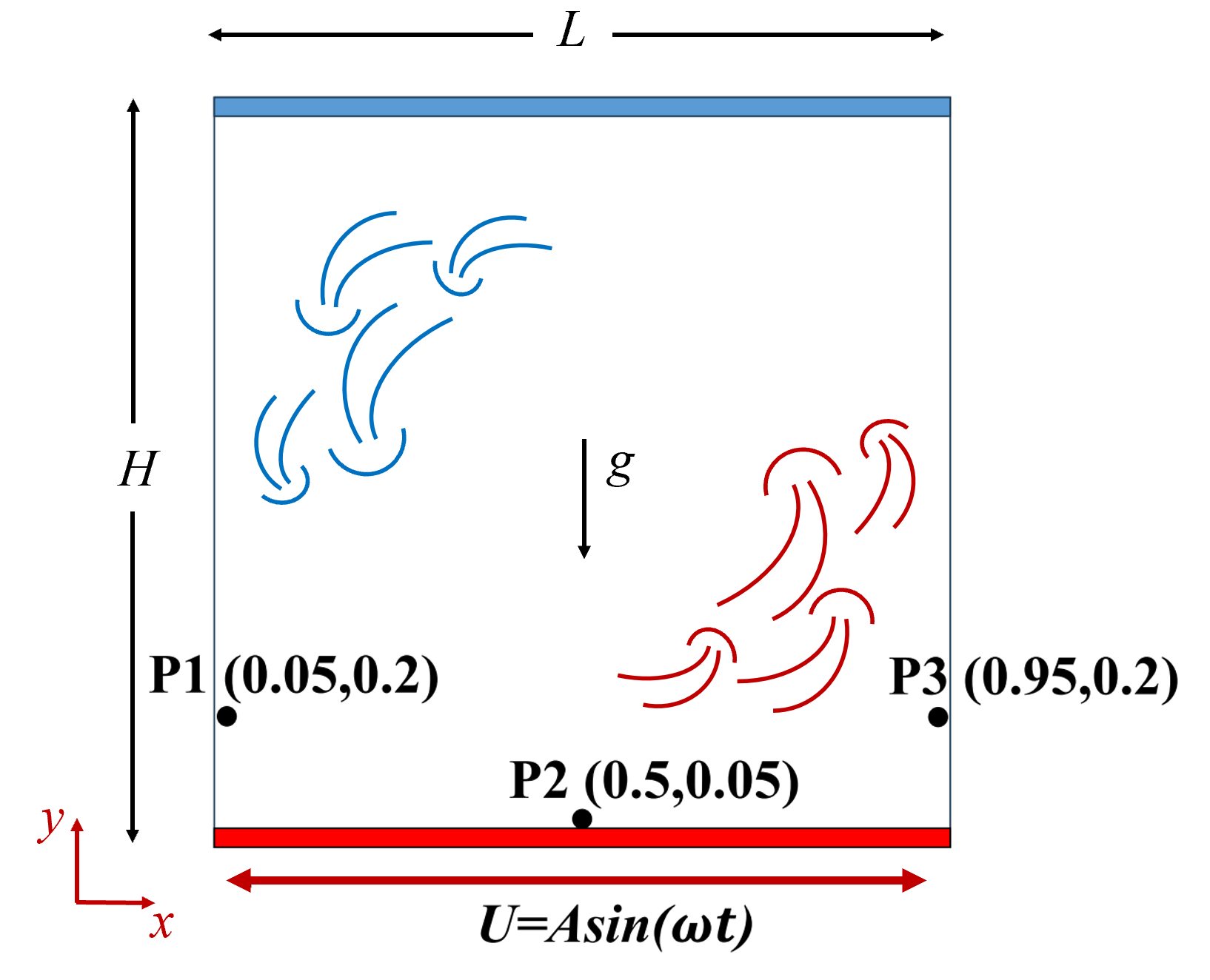}
        \put(2,78){(\textit{a})}  
    \end{overpic}
    \begin{overpic}[width=0.45\linewidth]{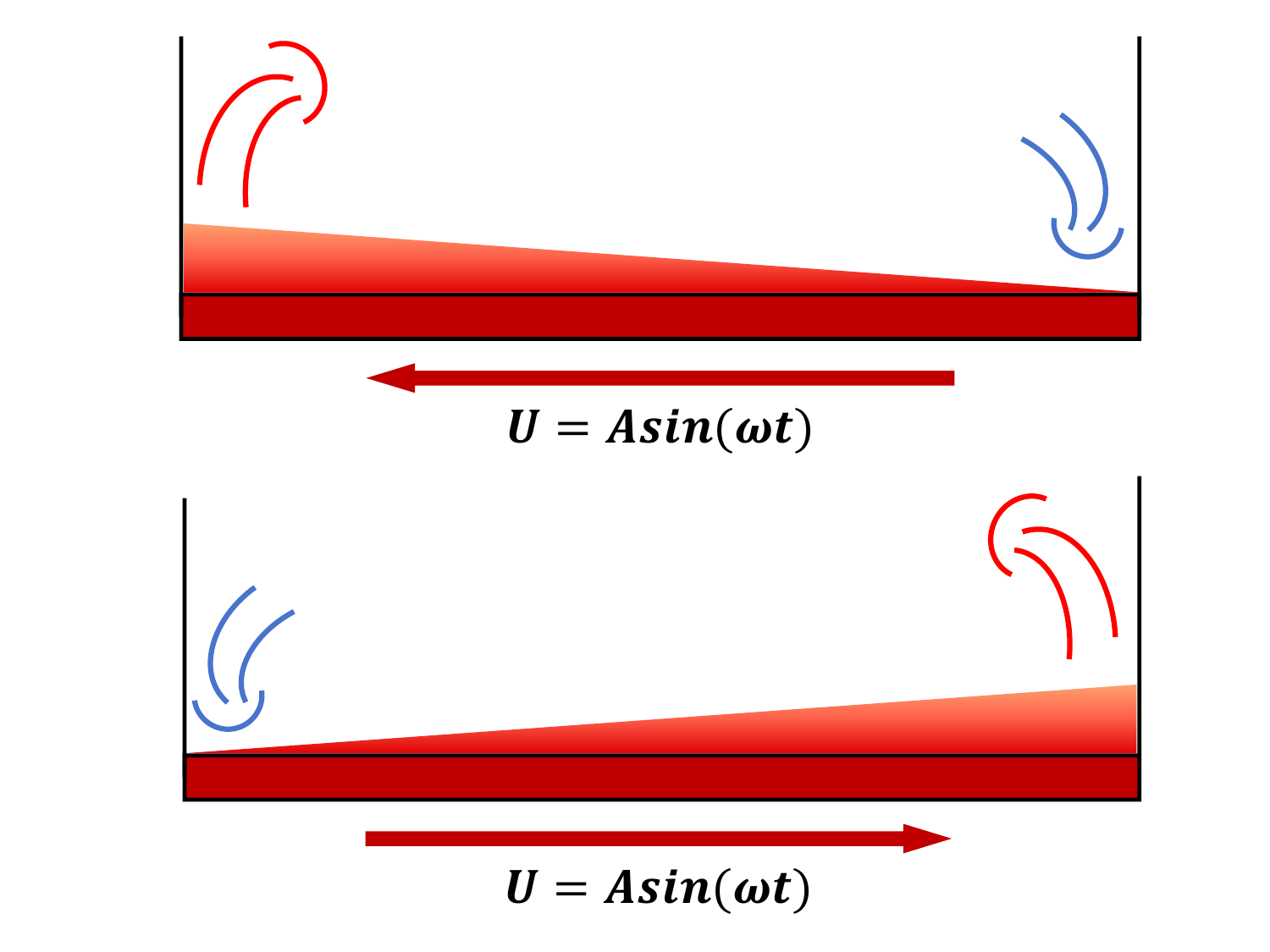}
        \put(2,86){(\textit{b})}
    \end{overpic}
    \caption{
    Schematic diagram of the wall-oscillation-controlled Rayleigh--B\'{e}nard convection system.
    The bottom plate oscillates horizontally with velocity $u(x,0,t)=A\sin(\omega t)$, while the top plate and sidewalls remain stationary.
    The vertical velocity $v$ at the positions of $P1(x=0.05, y=0.2)$ and $P3 (x=0.95, y=0.2)$, and the horizontal velocity $u$ at the position of $P2(x=0.5, y=0.05)$ are monitored to reveal the flow motions, which are related to the discussion of the  LSC reversals in the text.
    }
    \label{fig:Schematic diagram}
\end{figure}

For thermal boundary conditions, the horizontal plates are maintained at constant temperatures, while the sidewalls are adiabatic:
\begin{eqnarray}
\theta(x,0,t) = 0.5, \quad \theta(x,1,t) = -0.5, \label{eq:BC_th_1}\nonumber\\
\partial_x \theta(0,y,t) = 0, \quad \partial_x \theta(1,y,t) = 0. \label{eq:BC_th_2}
\end{eqnarray}
The velocity boundary conditions are no-slip on all surfaces. As shown in figure~\ref{fig:Schematic diagram}, the bottom plate is imposed with a time-periodic horizontal velocity corresponding to sinusoidal oscillation, while the top plate and sidewalls remain stationary:
\begin{eqnarray}
{u(x,0,t) = A\sin(\omega t)}, \qquad v(x,0,t) = 0, 
\end{eqnarray}
where $A$ represents the amplitude and $\omega=2\pi f$ with $f$ the oscillation frequency. Since we focus on the frequency effect, the magnitude $A$ is set as unity throughout this study.

The numerical simulations are performed using a finite difference method with second-order accuracy in both space and time. The governing equations are solved via a projection method. The pressure Poisson equation is efficiently solved using a fast Fourier transform in the horizontal ($x$) direction, while the tridiagonal systems in the vertical ($y$) direction are solved using a parallel diagonal dominant algorithm. The code has been extensively validated in our earlier work \citep{he2023scaling, He2024turbulent}. Here, the simulations are conducted at a fixed Prandtl number $Pr = 4.3$ with $Ra \in [5 \times 10^6, 1 \times 10^8]$, and the aspect ratio $\Gamma = L/H=1$. 
The global heat transport is quantified by the Nusselt number, computed from the heat flux as: 
$Nu = \sqrt{Ra Pr} \langle v \theta \rangle_{V,t} - \langle \partial_y \theta \rangle_{V,t}$, 
where $\langle \cdot \rangle_{V,t}$ denotes the temporal and spatial average over the entire system. 

The grid resolution is carefully chosen to resolve the smallest dynamical scales, namely the Kolmogorov scale $\eta_K$ and the Batchelor scale $\eta_B$. 
The Kolmogorov length scale is estimated as $\eta_K = (\nu^3 / \langle \varepsilon_u \rangle_{V,t} )^{1/4}$, and the Batchelor scale as $\eta_B = \eta_K Pr^{-1/2}$ \citep{xu2023wall}, where $\langle \varepsilon_u \rangle_{V,t}$ is the temporally and spatially averaged kinetic energy dissipation rate. 
The time step is constrained by the Kolmogorov time scale $\tau_\eta = \sqrt{\nu / \langle \varepsilon_u \rangle_{V,t}}$ to ensure temporal resolution. 
Our grid spacing $\Delta_g$ satisfies $(\Delta_g)_{\text{max}} / \eta_K < 1$ and $(\Delta_g)_{\text{max}} / \eta_B < 1$ for all cases, guaranteeing sufficient spatial resolution even under strong wall shear. 
All key simulation parameters, including $Ra$ frequency $f$, and grid resolution, are listed in Appendix~\ref{app:param}.
Statistical convergence was confirmed by comparing $Nu$ across successive intervals of at least $600 t_f$, with relative variations remaining below 1\%. Additionally, the discrepancy between $Nu$ calculated from heat flux and thermal dissipation was consistently less than 1\%, ensuring the reliability of our results.

\section{Optimal heat enhancement at $Ra=10^7$}
\label{sec3}

\subsection{Nusselt number variation}

\begin{figure}
   \centering
   \begin{overpic}[width=0.70\textwidth]{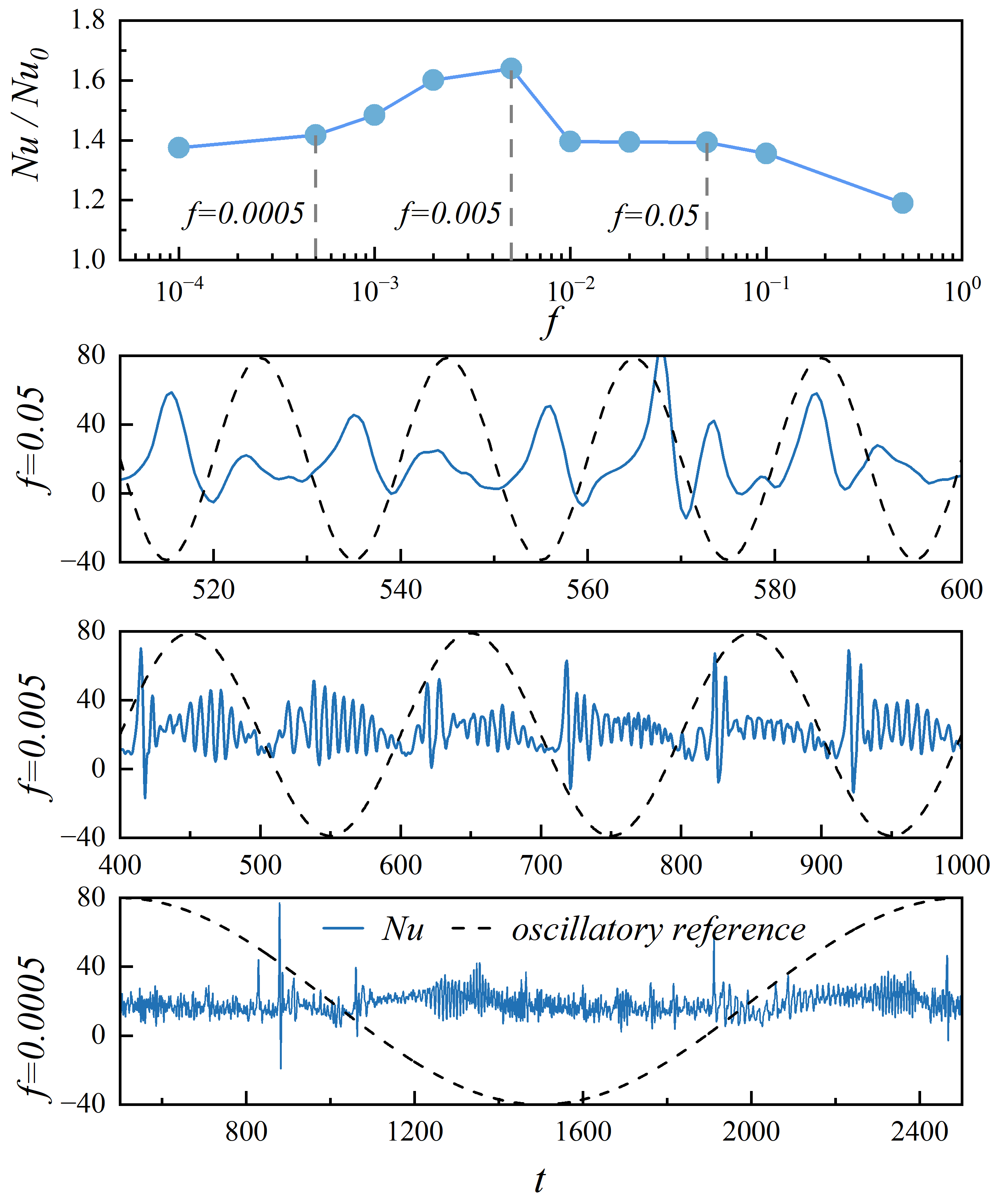}
      \put(2,100){(\textit{a})} 
     \put(2,70){(\textit{b})}
     \put(2,48){(\textit{c})}
     \put(2,26){(\textit{d})}
   \end{overpic}
\caption{Frequency dependence of (a) the normalized Nusselt number $Nu(f)/Nu_0$ for the case $Ra = 1\times10^{7}$ at $Pr = 4.3$, where $Nu_0$ denotes the value in the uncontrolled case.
Temporal evolution of the global Nusselt number under different oscillation frequencies:
(b) $f=0.05$ (high frequency), 
(c) $f=0.005$ (optimal frequency), and 
(d) $f=0.0005$ (low frequency). 
The dashed vertical lines in (a) mark the frequencies chosen for analysis in Section ~\ref{sec3}. The dashed lines in (b-d) represent the phase of the wall oscillation.}
\label{fig:Nu_inst}
\end{figure}
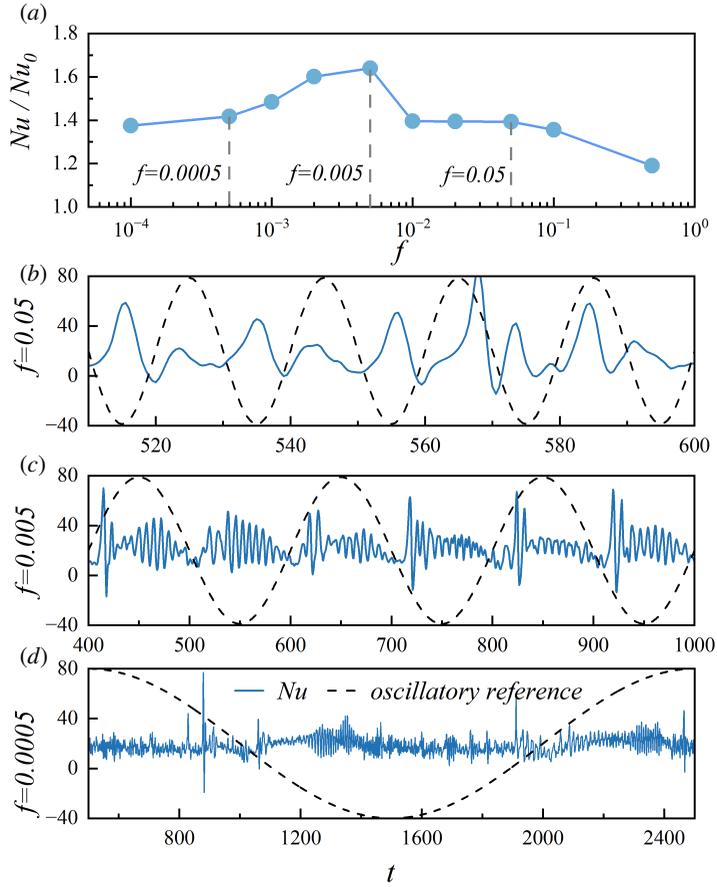

To quantify the heat-transfer enhancement, the normalized Nusselt number $Nu/Nu_0$, where $Nu_0$ denotes the Nusselt number of the uncontrolled case, exhibits three distinct frequency regimes in figure~\ref{fig:Nu_inst}(a): a low-frequency ascent ($f < 10^{-3}$) where $Nu/Nu_0$ increases monotonically; an intermediate-frequency regime ($10^{-3} \lesssim f \lesssim 2\times10^{-2}$) featuring a pronounced peak corresponding to the maximum enhancement; and a high-frequency decay ($f \gtrsim 2\times10^{-2}$) where the enhancement progressively weakens. These results demonstrate that $Ra = 1\times10^{7}$ exhibits a clear optimal forcing frequency at $f_{\text{opt}} = 0.005$, with $Nu$ increasing by over 60\%.

The temporal evolution of the instantaneous Nusselt number \(Nu(t)\) provides further insight into the frequency-dependent response of the system. Figures~\ref{fig:Nu_inst}(b–d) show \(Nu(t)\) for three representative forcing frequencies: high frequency \(f = 0.05\), optimal frequency \(f_{\mathrm{opt}} = 0.005\), and low frequency \(f = 0.0005\), respectively. Also included is the phase variation of the wall oscillation velocity, indicated by the dashed black line, whose correlations with $Nu$ are explained as follows.

At the high frequency \(f = 0.05\) (figure~\ref{fig:Nu_inst}b), the oscillation period is \(t_{\mathrm{osc}} = 20\) (after normalization by the free-fall time $t_f$). The instantaneous Nusselt number exhibits periodic variation, but with a frequency larger than that of the wall oscillation. One can see that for $510<t<520$, $Nu$ and the oscillatory velocity are negatively correlated; but for $580<t<590$, they become positively correlated. In contrast, at the optimal frequency \(f_{\mathrm{opt}} = 0.005\) (or the oscillation period \(t_{\mathrm{osc}} = 200\)), \(Nu(t)\) exhibits pronounced periodic peaks, whose envelope is synchronized well with the wall motion (figure~\ref{fig:Nu_inst}c). 
At the low frequency \(f = 0.0005\) (with period \(t_{\mathrm{osc}} = 2000\)), $Nu$ becomes more intermittent and almost has no correlation with the wall oscillation (figure~\ref{fig:Nu_inst}d). 
Also, comparing $f=0.05$ and $f=0.0005$, although they lead to a similar $Nu$ increase of approximately 40\%, the time responses of $Nu$ variation are quite different. To elucidate the physical mechanisms behind the responses of wall oscillations, the boundary layer structures and the large-scale circulations are monitored for different oscillation periods, explained now.





%

\subsection{Boundary layer behavior and LSC reversal}
\label{sec:lsc}

\begin{figure}
    \centering
    \begin{minipage}{1.0\linewidth}
   \centering
   \begin{overpic}[width=0.7\textwidth]{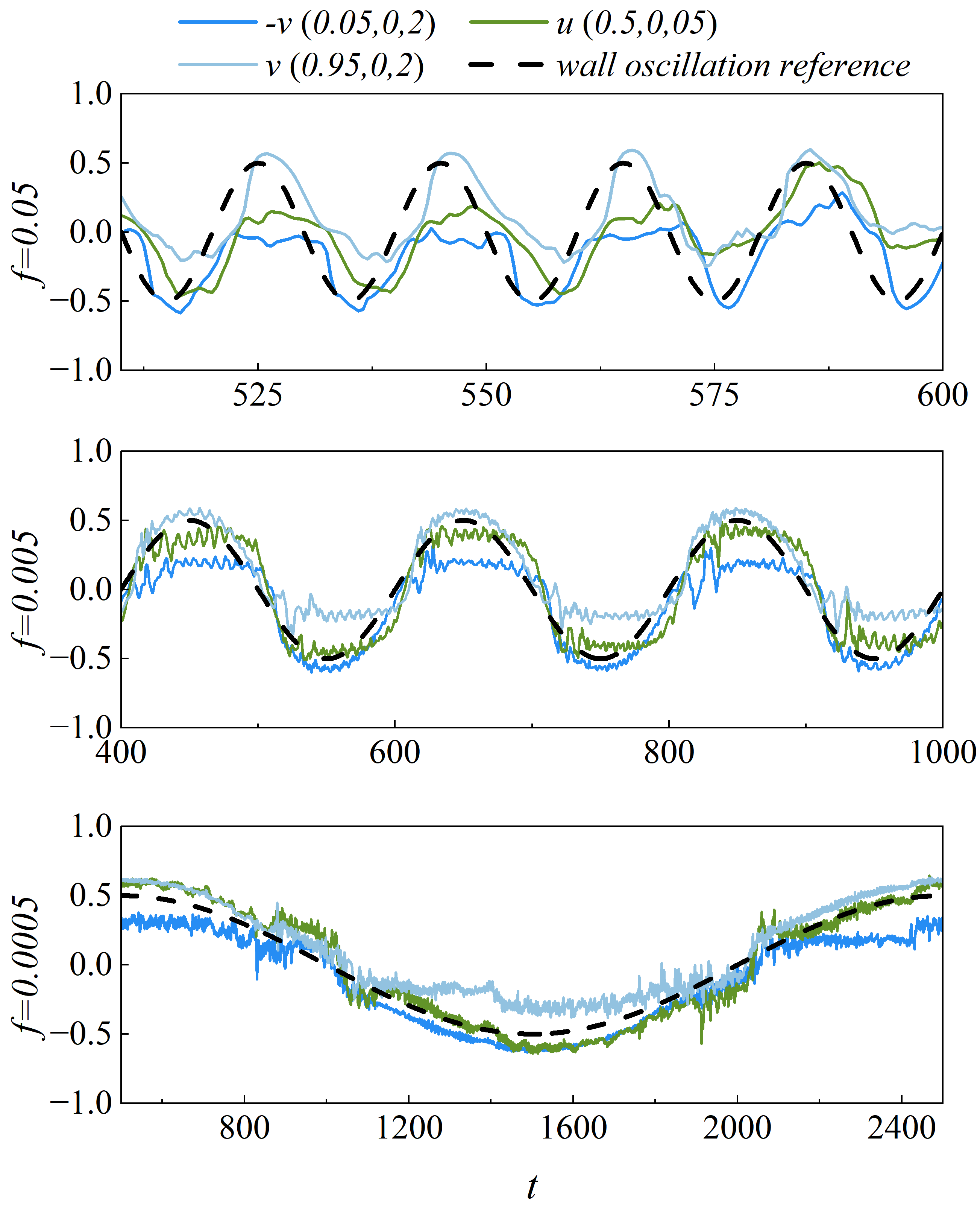}
     \put(3,95){(\textit{a})}
     \put(3,65){(\textit{b})}
     \put(3,35){(\textit{c})}

   \end{overpic}
  \end{minipage}
\caption{
Time series of velocity components at three monitoring points $P1(0.05,0.2)$, $P2(0.5,0.05)$ and $P3(0.95,0.2)$ (as shown in figure~\ref{fig:Schematic diagram}) for $Ra=1\times10^{7}$ at different oscillation frequencies. Rows from top to bottom correspond to $f=0.05$ (high frequency), $f=0.005$ (optimal frequency) and $f=0.0005$ (low frequency). Note that magnitude of wall oscillation velocity has been shifted to 0.5 for clarity.
}
\label{fig:probe_omega_1}
\end{figure}

\begin{figure}
    \centering
    \begin{minipage}{1.0\linewidth}
   \centering
   \begin{overpic}[width=0.7\textwidth]{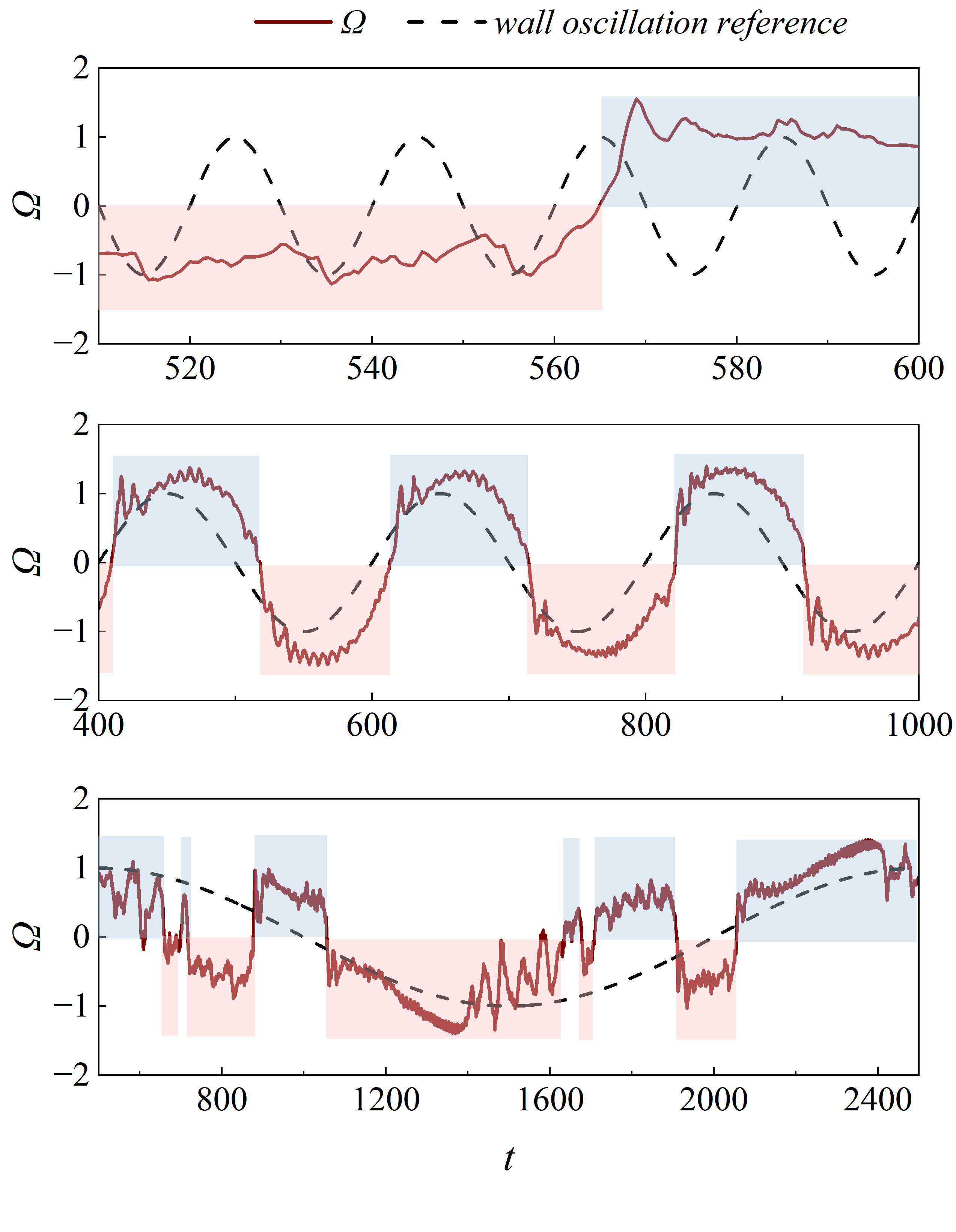}
     \put(0,95){(\textit{a})}
     \put(0,65){(\textit{b})}
     \put(0,35){(\textit{c})}

   \end{overpic}
  \end{minipage}
\caption{
Global angular momentum $\Omega(t)$ calculated using Eq.~(\ref{eq:omega}) for $Ra=1\times10^{7}$ at different oscillation frequencies. From top to bottom: $f=0.05$ (high frequency), $f=0.005$ (optimal frequency) and $f=0.0005$ (low frequency). The time series reveal the frequency-dependent response of the large-scale circulation to bottom-plate oscillation, with complete reversals occurring synchronously at the optimal frequency.
In each panel, the red region marks the interval when $\Omega<0$, while blue for $\Omega>0$.}
\label{fig:probe_omega_2}
\end{figure}

The response of the boundary layer to bottom-wall oscillation is examined using velocity probes located at $P1(0.05,0.2)$, $P2(0.5,0.05)$, and $P3(0.95,0.2)$ (figure~\ref{fig:Schematic diagram}a). As shown in figure~\ref{fig:probe_omega_1}, all monitored velocities exhibit periodic fluctuations that faithfully mirror the driving oscillation, suggesting that the near-wall flow responds effectively to the plate motion regardless of frequency. Specifically, at the high frequency of $f = 0.05$ (figure~\ref{fig:probe_omega_1}a), the vertical velocities near the sidewalls show alternating peaks with a magnitude of approximately $0.6$, indicating that the oscillating plate periodically squeezes fluid towards the sidewalls, while the horizontal velocity near the bottom center reaches peaks of about $0.4$, all adapting well to the imposed oscillation. At the optimal frequency $f_{\mathrm{opt}} = 0.005$ (figure~\ref{fig:probe_omega_1}b), a highly coherent response emerges, all velocities follow the wall oscillation with near-perfect synchrony. Even at the lowest frequency $f = 0.0005$ (figure~\ref{fig:probe_omega_1}c), the velocity signals follow the slow plate oscillation. 

Therefore, the consistent phase variation across all frequencies demonstrates that pointwise velocity measurements near the boundaries primarily capture the local kinematic response to the driving, and thus cannot by themselves reveal the frequency-dependent impact on global heat transport. A more comprehensive analysis, considering the spatial structure and phase relationships of the flow field, is therefore necessary to understand the enhancement of heat transfer at the optimal frequency.


We thus turn to focus on the orientation of the LSC. According to \cite{sugiyama2010flow, wang2018flow, zhang2020controlling}, the LSC orientation is characterized by the global angular momentum, defined as
\begin{equation} \label{eq:omega}
\Omega(t) = \left< \frac{v(x, y, t)}{x - 0.5} - \frac{u(x, y, t)}{y - 0.5} \right>_{V}, 
\end{equation}
where $\langle \cdot \rangle_{V}$ denotes the volume average over the entire system. Since $Nu$ is primarily governed by the convection of plumes, which are themselves driven by LSC, we expect that the variation of $\Omega(t)$ offers a useful means of detecting the efficiency of different wall oscillations

The time series of $\Omega(t)$ for the above three frequencies are presented in figure~\ref{fig:probe_omega_2}. At high frequency $f = 0.05$ (figure~\ref{fig:probe_omega_2}a), the global angular momentum reveals that a complete reversal of the LSC does not follow the wall oscillation period. In fact, the reversal occurs every several oscillation cycles, and the plate motion and LSC are not synchronous. This results in the intermittent nature of the heat-transfer observed in figure~\ref{fig:Nu_inst}(b), where the instantaneous Nusselt number $Nu(t)$ shows localized peaks when the wall motion aligns with the LSC, and dips below the average when they are opposed. At the optimal frequency $f_{\mathrm{opt}} = 0.005$ (figure~\ref{fig:probe_omega_2}b), the angular momentum displays six complete and clean reversals within the sampling window ($400t_f\leq t\leq1000t_f$), the same as the bottom-plate velocity that changes direction. These reversals occur approximately every $100t_f$, which is exactly half the oscillation period. This shows that the LSC reverses twice during each complete cycle of the bottom plate, locking itself in perfect sync with the external forcing. This perfectly synchronized reversal is the dynamical mechanism behind the sharp periodic peaks of $Nu(t)$ seen in figure~\ref{fig:Nu_inst}(c), demonstrating how the forcing efficiently extracts and ejects thermal plumes in the boundary (figure~\ref{fig:probe_omega_1}b) to maximize vertical heat transport. At the lowest frequency $f = 0.0005$ (figure~\ref{fig:probe_omega_2}c), the angular momentum signal exhibits frequent sign changes, with at least six zero-crossings occurring within a wall oscillation period, thus highlighting the difference compared to other oscillation frequencies. 

Therefore, from figure~\ref{fig:probe_omega_2}(a) to figure~\ref{fig:probe_omega_2}(c), the recovery in the $\Omega$-sign distinguishes the control efficiency. That is, at the optimal control frequency $f=f_{\mathrm{opt}}$, the sign of $\Omega$ changes exactly after the wall oscillation. In contrast, for $f>f_{\mathrm{opt}}$, the $\Omega$-sign recovers much slower than the oscillation; for $f<f_{\mathrm{opt}}$, the $\Omega$-sign changes more frequently than the wall oscillation. Below, we address two questions: how the flow structures respond to wall oscillation (\S3.3), and whether such a phase-locking mechanism between  $\Omega$ and $f$ persists for other values of $Ra$ (\S4).

\subsection{Fourier mode analysis}

The angular momentum analysis in figure~\ref{fig:probe_omega_2} reveals  different responses of the large-scale circulation to the oscillatory forcing across the three frequency regimes. To elucidate the underlying flow structures responsible for these distinct behaviors and to understand how they govern heat-transfer efficiency, we employ Fourier mode decomposition (FMD) \citep{chandra2011dynamics,petschel2011statistical,   wagner2013aspect,Gao2024JFM} to characterize the evolution of coherent flow modes under different forcing frequencies.




\begin{figure}
    \centering
    \begin{minipage}{1.0\linewidth}
   \centering
   \begin{overpic}[width=0.95\textwidth]{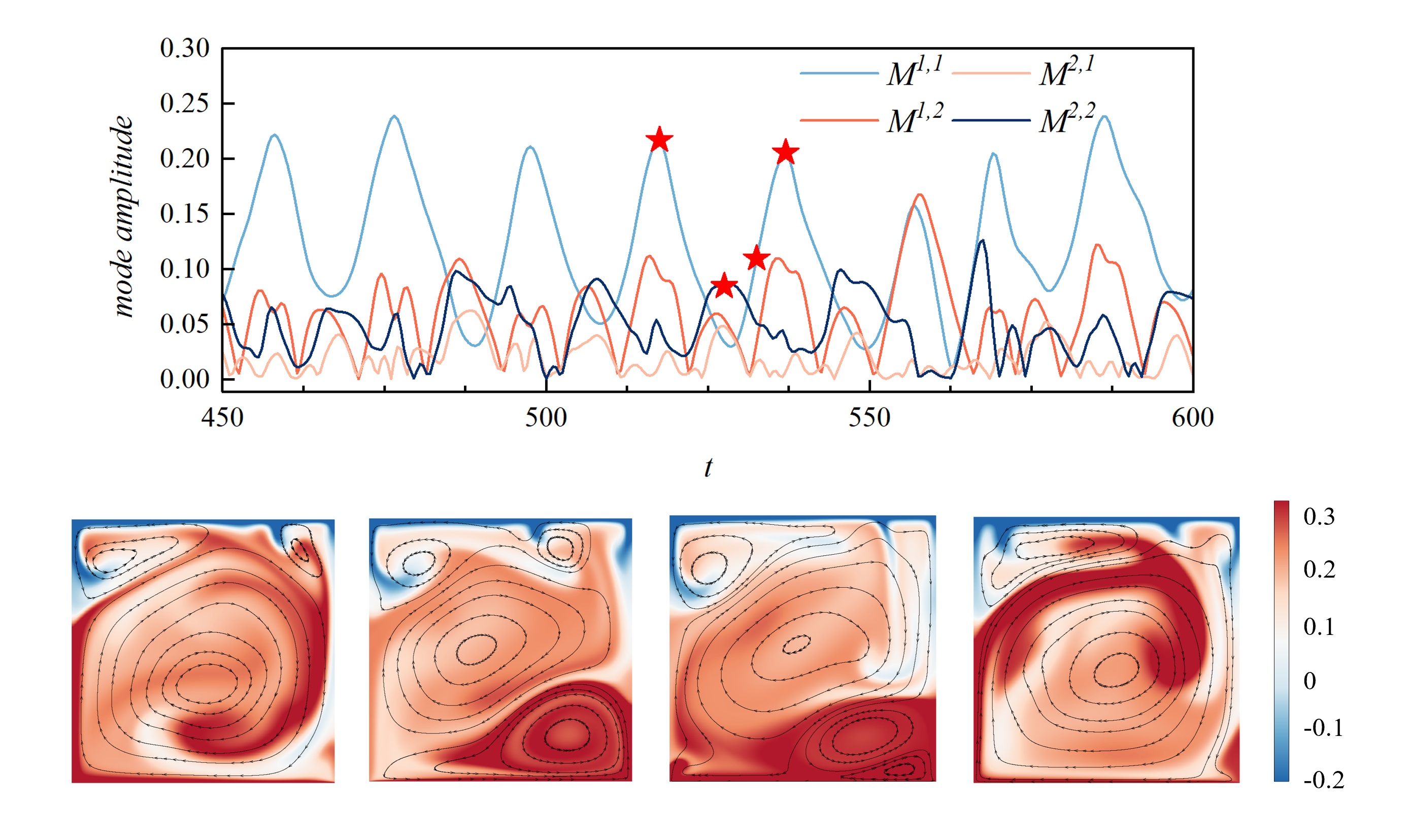}
     \put(7,57){(\textit{a})}
     \put(3,24){(\textit{b})}
     \put(24,24){(\textit{c})}
     \put(45,24){(\textit{d})}
     \put(66,24){(\textit{e})}
   \end{overpic}
  \end{minipage}
\caption{
Fourier-mode evolution and associated flow-structure transition at $Ra=1\times10^7$ and high forcing frequency $f=0.05$. 
Panel (a) shows the temporal evolution of the dominant Fourier mode amplitudes, where the red stars mark the time instants corresponding to the instantaneous flow fields shown in panels (b–e). 
Panels (b–e) display instantaneous temperature fields with velocity streamlines, representing different stages of a failed LSC reversal: 
(b) a stable LSC-dominated state, (c) the emergence of corner vortices, (d) deformation of the corner vortex induced by opposite wall motion, and (e) recovery to the original LSC configuration. 
}
\label{fig:fmd_high}
\end{figure}

\begin{figure}
    \centering
    \begin{minipage}{1.0\linewidth}
   \centering
   \begin{overpic}[width=0.85\textwidth]{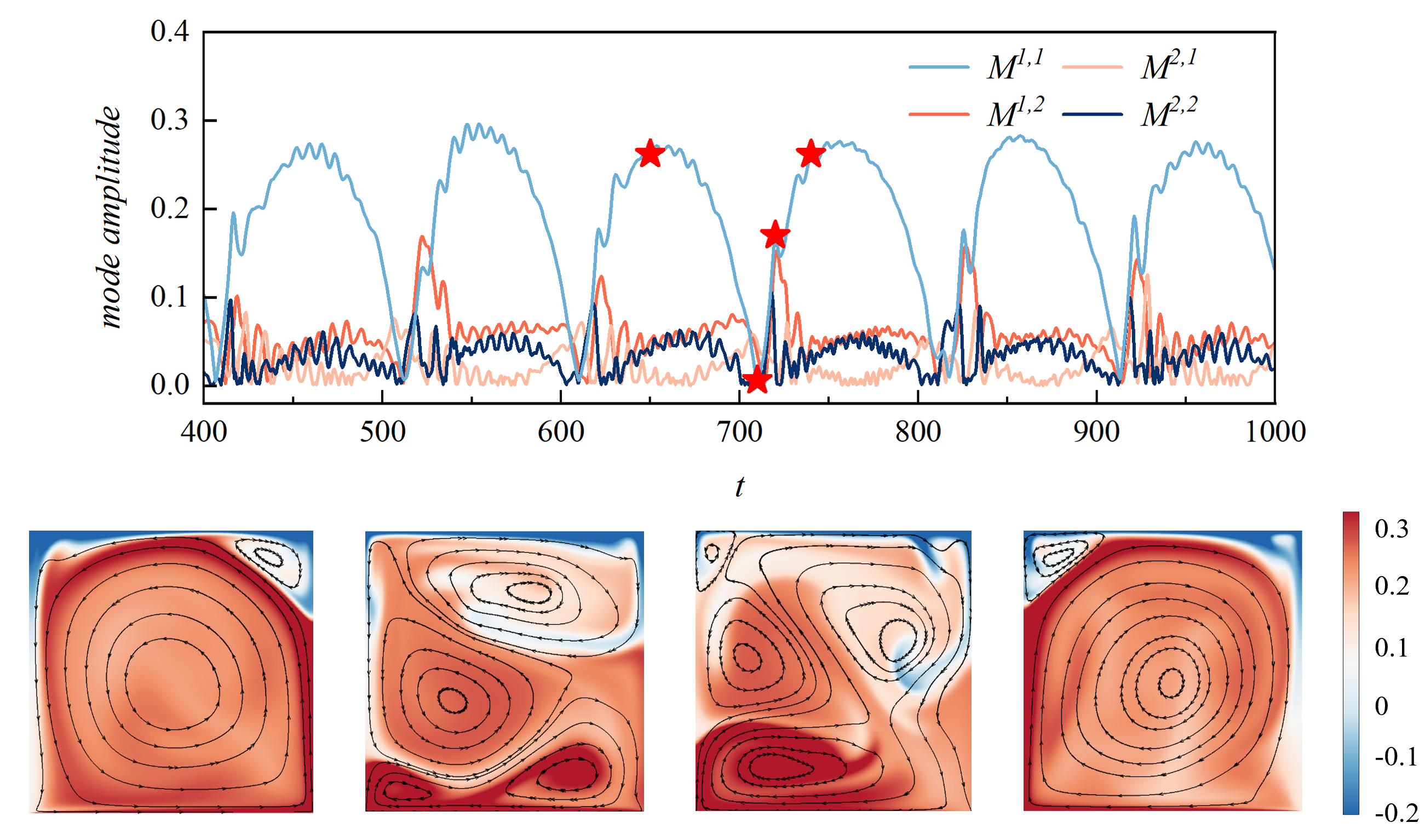}
     \put(5,57){(\textit{a})}
     \put(0,23){(\textit{b})}
     \put(24,23){(\textit{c})}
     \put(47,23){(\textit{d})}
     \put(70,23){(\textit{e})}
   \end{overpic}
  \end{minipage}
\caption{
Fourier-mode evolution and flow-structure transition at optimal frequency $f_{\text{opt}}=0.005$. 
Panel (a) shows the temporal evolution of the dominant Fourier mode amplitudes, where the red stars mark the time instants corresponding to the instantaneous flow fields shown in panels (b–e). 
Panels (b–e) display instantaneous temperature fields with velocity streamlines, representing different stages of a complete LSC reversal: 
(b) a stable LSC-dominated state corresponding to the $M^{1,1}$ mode, (c) emergence of the $M^{1,2}$ mode as two vertically stacked vortices develop and push the original LSC aside, (d) connection of the upper and lower vortices to form a large-scale structure that further displaces the initial LSC, and (e) consolidation of the new dominant vortex, establishing a fully reversed LSC now dominated again by the $M^{1,1}$ mode. 
}

\label{fig:fmd_opt}
\end{figure}

While details on FMD are presented in Appendix~\ref{app:FMD}, the results are explained below with reference to the temporal evolution of Fourier modes and the corresponding flow structures. Figure~\ref{fig:fmd_high} presents the FMD analysis for the high-frequency case $f=0.05$, where the oscillatory forcing is too rapid to permit a complete flow reversal, as previously noted in Figure~\ref{fig:probe_omega_2} (a). 
Panel (a) of Figure~\ref{fig:fmd_high} shows the temporal evolution of the dominant Fourier mode amplitudes. The $M^{1,1}$ mode , a single-roll mode representing the basic LSC structure remains dominant throughout, with its oscillation frequency matching the bottom-plate forcing frequency. However, the higher-order modes, i.e. $M^{1,2}$ (double-roll vertical mode), $M^{2,1}$ (double-roll horizontal mode), $M^{2,2}$ (quadrupole mode), exhibit irregular and non-periodic fluctuations, reflecting the disorganized response of secondary structures to rapid forcing. 
In Figure~\ref{fig:fmd_high}(b), a stable clockwise LSC is observed while the bottom plate moves leftward. 
As the plate motion reverses to the right, a corner roll develops in the lower-right region, as shown in Figure~\ref{fig:fmd_high}(c). 
However, due to the short oscillation period, this corner roll does not have sufficient time to grow before the plate motion switches back to the left. Consequently, its development is suppressed, and the accumulated hot fluid is advected leftward, forming a strong thermal plume (Figure~\ref{fig:fmd_high}d). 
The flow then recovers to its original LSC configuration (Figure~\ref{fig:fmd_high}e), illustrating a failed reversal. 
This sequence, also visualized in Movie 1, demonstrates that high-frequency forcing suppresses the growth of secondary vortices and inhibits complete LSC reorientation. This sequence illustrates the transient excitation and suppression of secondary vortical structures under high-frequency forcing, resulting in non-periodic and incomplete LSC reversals.

At the optimal frequency $f_{\text{opt}}=0.005$, the FMD analysis reveals a well-organized periodic behavior. 
Figure~\ref{fig:fmd_opt}(a) displays the temporal evolution of Fourier mode amplitudes, exhibiting a characteristic period of approximately $100t_f$, corresponding to half of the oscillation cycle. This periodicity is consistent with the probe velocities and angular momentum shown in Figures~\ref{fig:probe_omega_1}(b) and \ref{fig:probe_omega_2}(b). 
The $M^{1,1}$ mode remains dominant throughout most of the cycle, but during each reversal event, the amplitudes of the $M^{1,2}$ and $M^{2,1}$ modes increase significantly, reflecting the transient reorganization of the flow structure. 
The red stars in Figure~\ref{fig:fmd_opt}(a) mark the flow states depicted in panels (b–e) of Figure~\ref{fig:fmd_opt}, illustrating the correspondence between modal energy distribution and flow evolution during a complete LSC reversal.
Figure~\ref{fig:fmd_opt}(b) shows an initial state dominated by the $M^{1,1}$ mode, characterized by a counter-clockwise LSC with a small corner roll in the upper-right corner. 
As the oscillation proceeds, two small vortices emerge near the bottom plate, while the central vortex is gradually suppressed. The $M^{1,2}$ mode develops, as shown in Figure~\ref{fig:fmd_opt}(c), with the lower-left vortex strengthening and merging with the lower-right one to form a larger coherent structure. 
In Figure~\ref{fig:fmd_opt}(d), this newly formed vortex connects with the upper-right corner roll, giving rise to a large-scale vortex that expands throughout the cell. 
Driven by the bottom-plate motion, this large-scale vortex eventually establishes a new LSC with an opposite (clockwise) orientation, while the previous LSC is reduced to a corner roll near the upper-left corner, now again dominated by the $M^{1,1}$ mode (Figure~\ref{fig:fmd_opt}e). This sequence illustrates the transient excitation and reorganization of vortical structures under optimal-frequency forcing, leading to a periodic and complete LSC reversal.
This complete reversal process, lasting approximately $30t_f$ (about 30\% of each forcing period), is visualized in Movie 2. 
During the reversal, $Nu$ exhibits a rapid change, with periodic peaks in Figure~\ref{fig:Nu_inst}(b), consistent with previous studies \citep{wang2018flow, xu2020correlation} that associate such transients with flow reorganization. 
After reversal, the LSC is accelerated by the bottom plate shear, enhancing plume transport and elevating $Nu/Nu_0$.

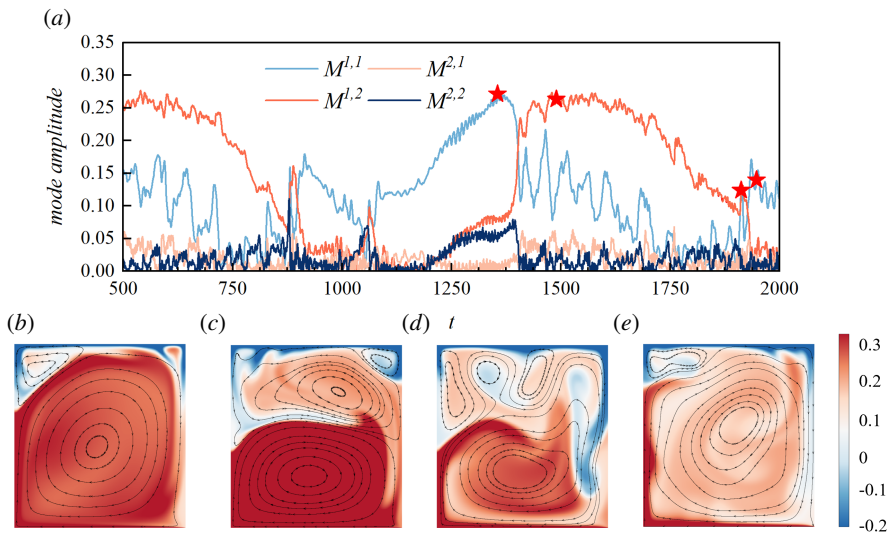
\begin{figure}
    \centering
    \begin{minipage}{1.0\linewidth}
   \centering
   \begin{overpic}[width=0.9\textwidth]{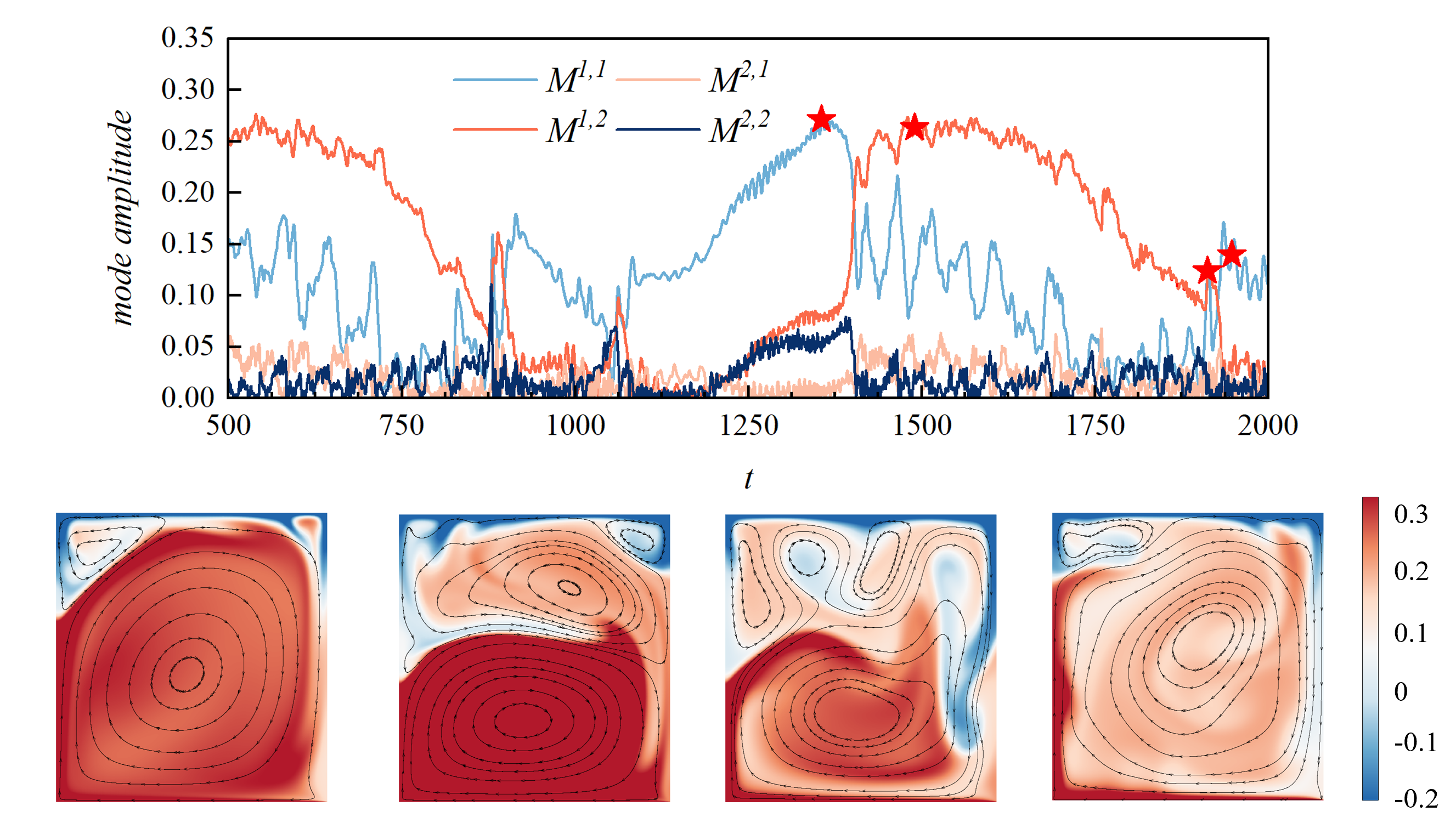}
     \put(7,57){(\textit{a})}
     \put(3,24){(\textit{b})}
     \put(24,24){(\textit{c})}
     \put(46,24){(\textit{d})}
     \put(69,24){(\textit{e})}
   \end{overpic}
  \end{minipage}
\caption{
Fourier-mode evolution and flow-structure transition at low frequency  $f=0.0005$. Panel (a) presents the temporal evolution of the dominant Fourier mode amplitudes, where the red stars mark the time instants corresponding, from left to right, to the instantaneous flow fields shown in panels (b–e).
Panels (b–e) show instantaneous temperature fields with velocity streamlines, illustrating the gradual evolution and eventual completion of an LSC reversal: 
(b) an LSC-dominated single-vortex state, (c) a fully developed double-roll structure, (d) the $\Omega$-transition phase where two vortices merge into one, and (e) the post-transition state with a new dominant LSC. 
}
\label{fig:fmd_low}
\end{figure}

At the low-frequency case $f=0.0005$, Figure~\ref{fig:fmd_low}(a) shows the temporal evolution of the dominant Fourier mode amplitudes. 
Unlike the optimal case, the $M^{1,1}$ and $M^{1,2}$ modes undergo multiple transitions within a single oscillation period, during which they persist as the dominant mode alternately. 
After each reversal, the $M^{1,1}$ mode persists for about $330t_f$, and then the flow mode transforms into the $M^{1,2}$ mode again under the strong shear induced by motion of the bottom plate. 
Figure~\ref{fig:fmd_low}(b) depicts a state dominated by the $M^{1,1}$ mode, where a distinct LSC occupies nearly the entire cell, accompanied by a corner roll in the upper-left corner. 
As high-speed flow impacts the top plate, fluid is deflected leftward, transferring energy into the corner roll. 
This intensification leads to a transition to a double-roll structure, as shown in Figure~\ref{fig:fmd_low}(c), where the $M^{1,2}$ mode becomes dominant. 
In this double-roll state, two vertically stacked vortices compete, causing the angular momentum variations observed in Figure~\ref{fig:probe_omega_2}(c). 
Before each reversal, as the bottom shear velocity decreases to zero, the flow mode switches back to $M^{1,1}$. 
This transition is illustrated in Figures~\ref{fig:fmd_low}(d) and \ref{fig:fmd_low}(e): a blue corner roll emerges in the upper-right corner and connects with the lower roll, squeezing the upper roll (Figure~\ref{fig:fmd_low}d), eventually developing into a new LSC dominated by the $M^{1,1}$ mode (Figure~\ref{fig:fmd_low}e). 
The dynamic evolution of these transitions is presented in Movie 3.

Comparing figure \ref{fig:fmd_low}(a) and figure~\ref{fig:Nu_inst}(c), it is notable that when the $M^{1,2}$ mode takes the dominance, $Nu(t)$ remains below its mean value, in contrast to the $M^{1,1}$ mode dominance during which $Nu(t)$ exceeds its mean value. The emergence of this less efficient mode $M^{1,2}$ explains the observed decrease in $Nu/Nu_0$ at low frequency $f=0.0005$, despite the intensified plume generation. This behavior is consistent with the result of \citet{xu2020correlation,xu2023wall}, who showed that the vertically stacked double-roll mode is inefficient for heat transfer on average. 

Together, the flow responses at different oscillation frequencies exhibit distinct behaviors.
At optimal frequency, the LSC reverses synchronously with the plate motion, leading to efficient plume transport and enhanced heat transfer. However, at low and high frequencies, either the formation of a vertically stacked double-roll structure or the occurrence of incomplete reversals disrupts the coherence of the LSC, resulting in reduced heat-transfer efficiency compared with the optimal case. In the following, we verify the phase-locking mechanism for optimal heat transfer at different $Ra$'s.

\section{Verifying the phase-locking mechanism at different \(Ra\)'s}

\begin{figure}
    \centering
    \begin{minipage}{1.0\linewidth}
   \centering
   \begin{overpic}[width=0.95\textwidth]{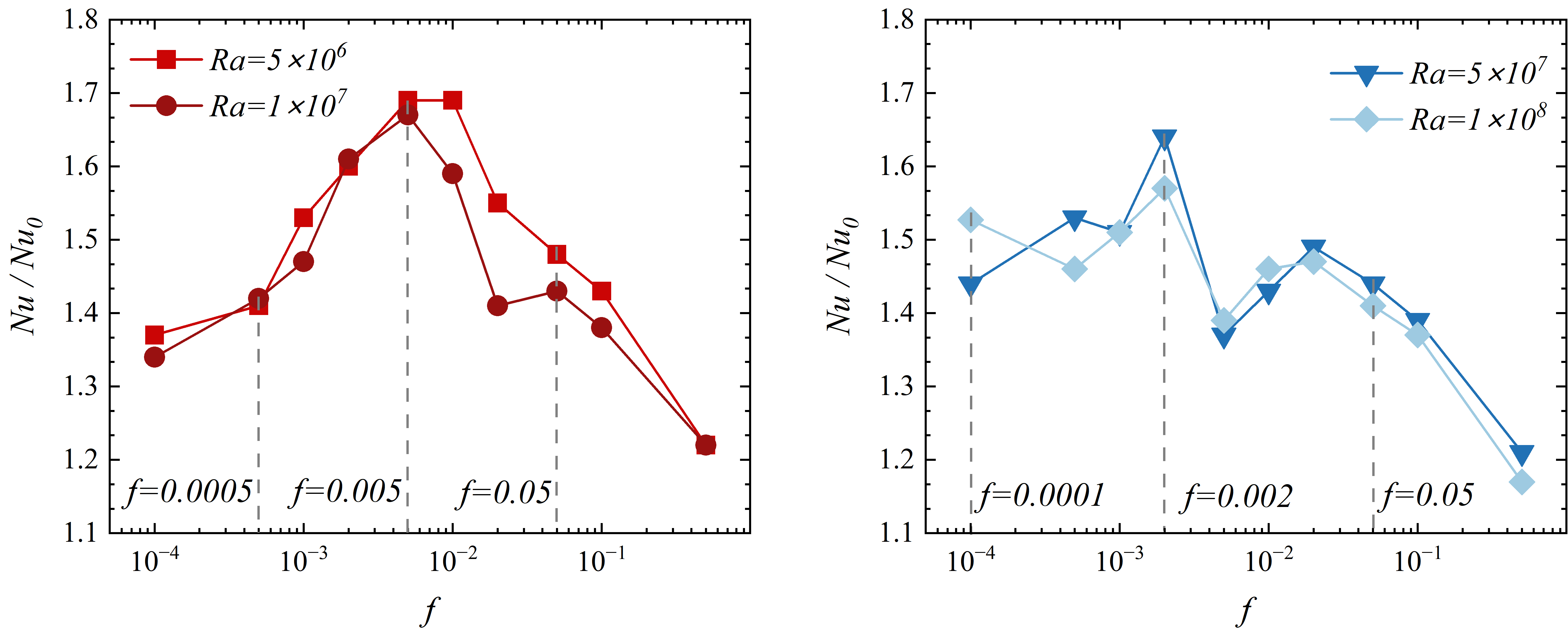}
     \put(3,41){(\textit{a})}
     \put(55,41){(\textit{b})}
   \end{overpic}
  \end{minipage}
\caption{Frequency dependence of the normalized Nusselt number $Nu(f)/Nu_0$. (a): $Ra=5\times10^{6}$ and $Ra=1\times10^{7}$; (b): $Ra=5\times10^{7}$ and $Ra=1\times10^{8}$. The dashed lines indicate the frequencies selected for analysis in figures~\ref{fig:probe_omega_5e6} and~\ref{fig:probe_omega_1e8}.}
\label{fig:Nu and Re}
\end{figure}

\begin{figure}
    \centering
    \begin{minipage}{1.0\linewidth}
   \centering
   \begin{overpic}[width=0.95\textwidth]{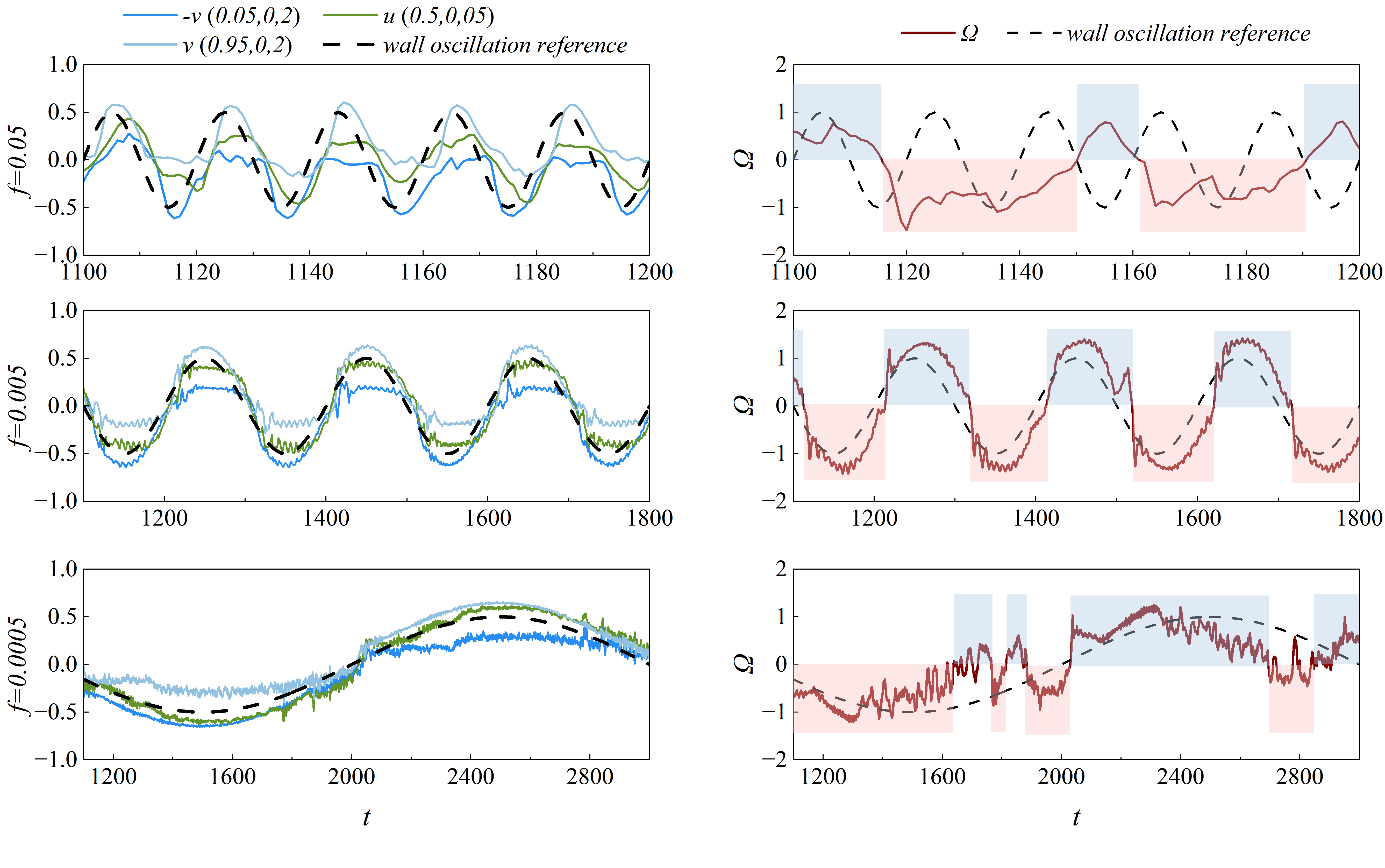}
     \put(0,58){(\textit{a})}
     \put(0,40){(\textit{b})}
     \put(0,22){(\textit{c})}
     \put(52,58){(\textit{d})}
     \put(52,40){(\textit{e})}
     \put(52,22){(\textit{f})}

   \end{overpic}
  \end{minipage}
\caption{
Velocity and angular momentum dynamics at $Ra=5\times10^{6}$:
First column (a-c): Time series of velocity components at three monitoring points $(0.05,0.2)$, $(0.5,0.05)$ and $(0.95,0.2)$, as shown in figure~\ref{fig:Schematic diagram}. Note that magnitude of wall oscillation velocity has been shifted to 0.5 for clarity. Second column (d-f): Global angular momentum $\Omega(t)$ calculated using Eq.~(\ref{eq:omega}).
Rows from top to bottom correspond to $f = 0.05$, $f = 0.005$ and $f = 0.0005$.
}
\label{fig:probe_omega_5e6}
\end{figure}

\begin{figure}
    \centering
    \begin{minipage}{1.0\linewidth}
   \centering
   \begin{overpic}[width=0.95\textwidth]{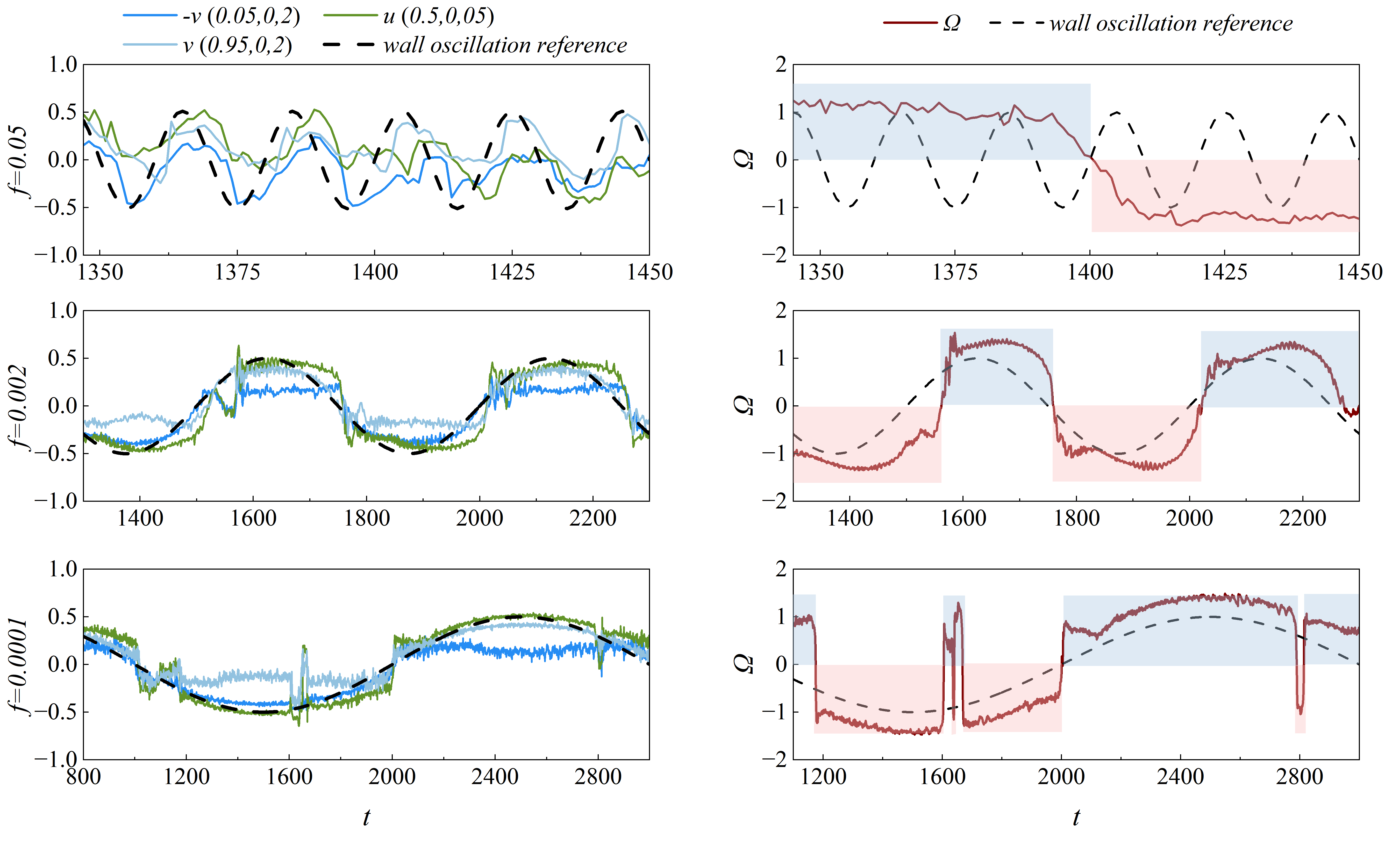}
     \put(0,58){(\textit{a})}
     \put(0,40){(\textit{b})}
     \put(0,22){(\textit{c})}
     \put(52,58){(\textit{d})}
     \put(52,40){(\textit{e})}
     \put(52,22){(\textit{f})}

   \end{overpic}
  \end{minipage}
\caption{
Velocity and angular momentum dynamics at $Ra=1\times10^{8}$:
First column (a-c): Time series of velocity components at three monitoring points $(0.05,0.2)$, $(0.5,0.05)$ and $(0.95,0.2)$, as shown in figure~\ref{fig:Schematic diagram}.Note that magnitude of wall oscillation velocity has been shifted to 0.5 for clarity. Second column (d-f): Global angular momentum $\Omega(t)$ calculated using Eq.~(\ref{eq:omega}).
Rows from top to bottom:
$f=0.05$;
$f=0.002$;
$f=0.0001$.}
\label{fig:probe_omega_1e8}
\end{figure}

The frequency-dependent heat-transfer enhancement observed at \(Ra = 1\times 10^{7}\) persists across the full range of Rayleigh numbers investigated, as summarized in figure~\ref{fig:Nu and Re}. All cases exhibit a consistent qualitative trend that modest enhancement at low forcing frequencies, a pronounced maximum at an intermediate optimal frequency, and a gradual decay as the frequency becomes sufficiently high. Note that compared to the distinct unimodal variation of $Nu$ in figure~\ref{fig:Nu and Re}(a), for higher $Ra$($>10^7$), $Nu$ shows an abrupt shrinkage at $f=0.005$ in figure~\ref{fig:Nu and Re}(b). Although this difference, it is important to check whether the underlying mechanism governing the optimal response remains robust. As marked in figure~\ref{fig:Nu and Re}, we select again three typical frequencies to check the phase-locking mechanism, i.e., an optimal frequency leading to maximum heat transfer, a lower frequency, and a higher frequency with less efficient heat transfer.

Taking \(Ra = 5 \times 10^{6}\) and \(Ra = 1 \times 10^{8}\) for example, the flow dynamics are examined in figures~\ref{fig:probe_omega_5e6} and~\ref{fig:probe_omega_1e8}, respectively. For both Rayleigh numbers, the near-wall velocity signals (figures~\ref{fig:probe_omega_5e6}a--c and~\ref{fig:probe_omega_1e8}a--c) exhibit periodic fluctuations that faithfully mirror the driving oscillation, indicating effective near-wall response regardless of frequency. The angular momentum dynamics reveal consistent frequency-dependent behavior, though with quantitative differences reflecting the increased inertial resistance at higher \(Ra\). At the optimal frequency (figure~\ref{fig:probe_omega_5e6}e for \(Ra = 5 \times 10^{6}\) with \(f_{\mathrm{opt}} \approx 0.005\); figure~\ref{fig:probe_omega_1e8}e for \(Ra = 1 \times 10^{8}\) with \(f_{\mathrm{opt}} \approx 0.002\)), the LSC reverses synchronously twice per oscillation cycle, maximizing vertical heat transport. At high frequencies (figure~\ref{fig:probe_omega_5e6}d and~\ref{fig:probe_omega_1e8}d, \(f = 0.05\) for both), reversals become intermittent and incomplete, requiring multiple cycles to complete a single reorientation. At low frequencies (figure~\ref{fig:probe_omega_5e6}f for \(Ra = 5 \times 10^{6}\) with \(f = 0.0005\); figure~\ref{fig:probe_omega_1e8}f for \(Ra = 1 \times 10^{8}\) with \(f = 0.0001\)), the system exhibits multiple angular momentum sign changes within a single oscillation period. Notably, in figure~\ref{fig:probe_omega_1e8}f, the angular momentum follows the wall oscillation more closely compared to the lower-\(Ra\) case, with fewer sign changes and shorter transition durations. The better synchronization of the LSC with the wall oscillation suggests that the low-frequency control ($f=0.0001$) becomes more effective with increasing \(Ra\), consistent with a larger $Nu$ for \(Ra = 1 \times 10^{8}\) than \(Ra = 5 \times 10^{7}\) shown in figure~\ref{fig:Nu and Re}b. 

Finally, through the Fourier mode analysis, the dominant structures for these additional $Ra$'s cases are found to be similar to the case of $Ra=10^7$. That is, at the optimal frequency the single-roll mode ($M^{1,1}$) dominates the flow organization and synchronizes well with the wall oscillation, while the LSC reversal is incomplete for high frequency or loses its dominance compared to the double-roll structure ($M^{1,2}$) at low frequency. The dynamics of structures are similar to those in figures~\ref{fig:fmd_high}-\ref{fig:fmd_low}, therefore not repeated here.

\section{Conclusion and Discussion}

Direct numerical simulations of two-dimensional Rayleigh--Bénard convection with a horizontally oscillating bottom plate have been performed, showing a significant enhancement of heat transfer up to 60\% compared to the uncontrolled case. With oscillation frequencies spanning three decades ($f = 0.0001$ to $0.5$), a phase-locking mechanism is identified: at the optimal frequency $f = f_{\text{opt}}$, the response time of the large-scale circulation (LSC) locks precisely to the oscillation period of the wall. In contrast, the LSC reversal time becomes substantially longer when $f > f_{\text{opt}}$ and significantly shorter when $f < f_{\text{opt}}$. This frequency-locking mechanism persists throughout the investigated range of Rayleigh numbers ($Ra = 5 \times 10^{6}$ to $1 \times 10^{8}$) at $Pr = 4.3$. Meanwhile, velocity signals within the boundary layers are found to be unable to distinguish control efficiency across different frequencies, as they remain consistently synchronized with the wall oscillation. Moreover, it is shown that at optimal frequency, the single-roll
mode remains dominant throughout the cycle, whilst at higher frequencies, the LSC cannot follow the rapid
forcing which leads incomplete reversals, and at lower
frequencies, a double-roll structure emerges which reduces the heat-transfer efficiency.

The present investigation is limited to two-dimensional configurations at a fixed Prandtl number and moderate Rayleigh numbers. Whether the same mechanisms persist in fully three-dimensional turbulence, and how they are influenced by the oscillation amplitude, remain open questions that warrant further investigation.


\backsection[Acknowledgements]{We appreciate Dr. A. Xu from Northwest University of Technology for helpful discussions. X.C. acknowledges support from the National Key Research and
Development Program of China (2022YFF0610805), and the National Natural Science Foundation of China, grant nos.92252201.}

\backsection[Declaration of interests]{The  authors  report  no  conflict  of  interest.}


\appendix

\section{Simulation parameters}
\label{app:param}
This table~\ref{tbl:Num} summarizes the key simulation parameters used in this study, including the Rayleigh number $Ra$, oscillation frequency $f$, grid resolution, averaging time, and related parameter calculations. All examples meet the spatial and temporal resolution criteria required for sufficient resolution of the minimum dynamic scale, and the reported statistics were obtained after the system reached statistical steady state.

\begin{table}
    \centering
    \setlength{\tabcolsep}{8pt}
    \begin{tabular}{c ccc ccc ccc ccc}
    \hline
    & \multicolumn{2}{c}{$Ra=5\times10^6$} & \multicolumn{4}{c}{$Ra=1\times10^7$} & \multicolumn{2}{c}{$Ra=5\times10^7$} & \multicolumn{4}{c}{$Ra=1\times10^8$} \\
    $f$ & $t_{avg}$ & $Nu$ & & $t_{avg}$ & $Nu$ & & $t_{avg}$ & $Nu$ & & $t_{avg}$ & $Nu$ & \\
    \hline
    0     & 1200 & 10.38 & & 1200 & 13.11 & & 1200 & 20.75 & & 1200 & 25.90 & \\
    0.0001   & 2500 & 14.26 & & 2500 & 17.54 & & 10000 & 29.90 & & 10000 & 39.56 & \\
    0.0005   & 2000 & 14.66 & & 2000 & 18.64 & & 3000 & 31.68 & & 3000 & 37.96 & \\
    0.001    & 1600 & 15.87 & & 1600 & 19.23 & & 4000 & 31.38 & & 4000 & 39.30 & \\
    0.002    & 1200 & 16.56 & & 1200 & 21.05 & & 5000 & 34.12 & & 2500 & 40.73 & \\
    0.005    & 1200 & 17.57 & & 1200 & 21.86 & & 4000 & 28.34 & & 2000 & 36.14 & \\
    0.01     & 1200 & 17.53 & & 1200 & 20.90 & & 1200 & 29.63 & & 1200 & 37.87 & \\
    0.02     & 1200 & 16.09 & & 1200 & 18.48 & & 1200 & 30.97 & & 1200 & 38.12 & \\
    0.05     & 1200 & 15.41 & & 1200 & 18.72 & & 1200 & 29.90 & & 1200 & 36.53 & \\
    0.1      & 1200 & 14.86 & & 1200 & 18.12 & & 1200 & 28.92 & & 1200 & 35.57 & \\
    0.5      & 1200 & 12.67 & & 1200 & 15.95 & & 1200 & 25.05 & & 1200 & 30.26 & \\
    \hline
    \end{tabular}
    \caption{The settings of simulations in this work. Here, $t_{avg}$ is the statistical averaging time after the flow fully developed. The grid resolutions are $N_x \times N_y = 512 \times 648$ for $Ra = 5 \times 10^{6}$ and $1 \times 10^{7}$, and $N_x \times N_y = 1024 \times 1296$ for $Ra = 5 \times 10^{7}$ and $1 \times 10^{8}$.}
    \label{tbl:Num}
\end{table}

\section{Fourier mode decomposition (FMD)}
\label{app:FMD}

FMD provides a systematic framework for separating flow modes according to their spatial symmetries and scale , and is particularly useful for characterizing the evolution of coherent structures during LSC reversals \citep{chen2020reduced,zhao2022suppression}.

\begin{figure}
\centering
    \includegraphics[width=0.9\textwidth]{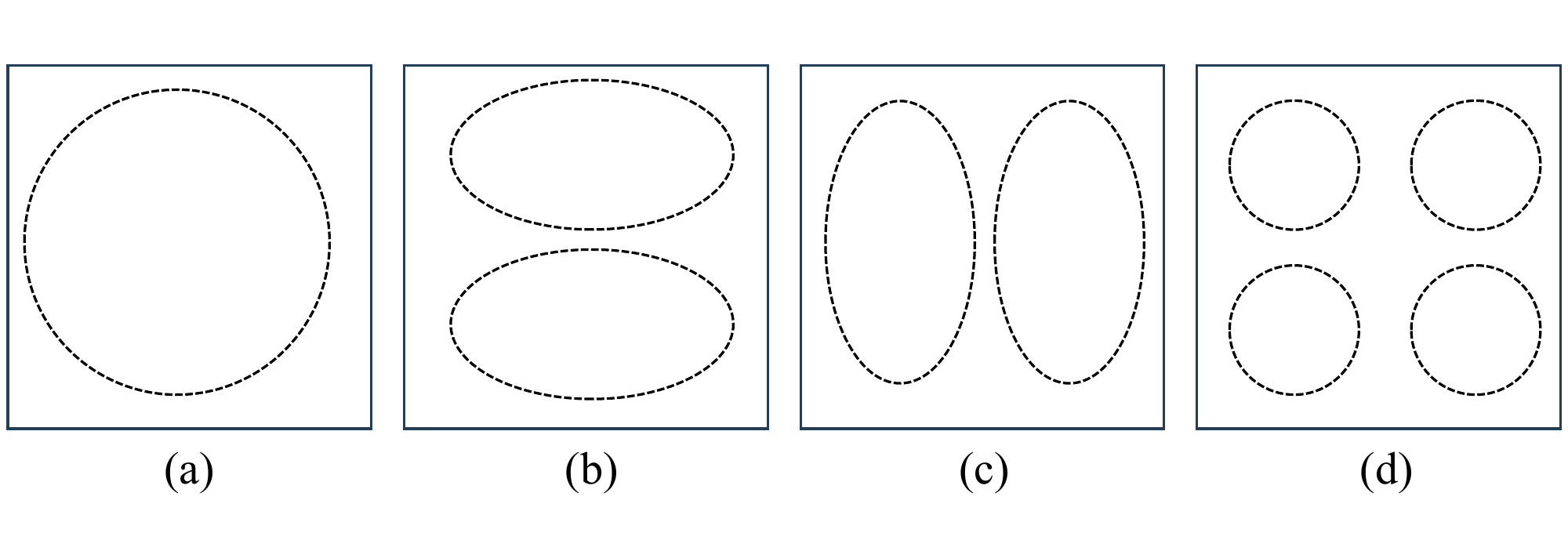}
\caption{
Schematic illustration of the FMD \citep{chandra2011dynamics,xu2020correlation}. 
(a) $M^{1,1}$: single-roll mode representing the basic LSC structure; 
(b) $M^{1,2}$: double-roll vertical mode; 
(c) $M^{2,1}$: double-roll horizontal mode; 
(d) $M^{2,2}$: quadrupole mode.
}
\label{fig:fourier_schematic}
\end{figure}

Specifically, the instantaneous horizontal and vertical velocity fields, $u(x,z,t)$ and $v(x,z,t)$, are expanded onto an orthogonal set of trigonometric basis functions defined over the rectangular computational domain. Each snapshot of the flow field is decomposed according to
\begin{eqnarray}
u(x,y,t) = \sum_{m,n} A_{m,n}^x(t) \, \hat{u}_{m,n}(x,y), \\
v(x,y,t) = \sum_{m,n} A_{m,n}^y(t) \, \hat{v}_{m,n}(x,y),
\end{eqnarray}
where the basis functions are defined as
\begin{eqnarray}
\hat{u}_{m,n} = 2 \sin(m\pi x)\cos(n\pi y), \\
\hat{v}_{m,n} = -2 \cos(m\pi x)\sin(n\pi y),
\end{eqnarray}
with $m$ and $n$ representing the horizontal and vertical mode numbers, respectively. The prefactor 2 ensures normalization consistency and energy conservation in the projection process. 

Following the projection procedure adopted in the previous studies \citep{xu2023wall, Gao2024JFM}, the time-dependent modal coefficients are evaluated by spatial projection of the instantaneous velocity components onto their corresponding basis functions:
\begin{eqnarray}
A_{m,n}^x(t) = \int_0^1 \int_0^1 u(x,y,t) \, \hat{u}_{m,n}(x,y) \, \mathrm{d}x \, \mathrm{d}y, \\
A_{m,n}^y(t) = \int_0^1 \int_0^1 v(x,y,t) \, \hat{v}_{m,n}(x,y) \, \mathrm{d}x \, \mathrm{d}y.
\end{eqnarray}
These coefficients quantify the instantaneous projection of the flow field onto each orthogonal spatial mode, effectively capturing the temporal evolution of their contributions to the overall flow dynamics.

The magnitude of each combined mode, denoted by $M_{m,n}(t)$, measures the instantaneous kinetic energy content associated with that mode and is computed as
\begin{equation}
M_{m,n}(t) = \sqrt{\left[A_{m,n}^x(t)\right]^2 + \left[A_{m,n}^y(t)\right]^2}.
\end{equation}
In the present analysis, we focus on the first four low-order Fourier modes, $M^{1,1}$, $M^{1,2}$, $M^{2,1}$, and $M^{2,2}$, which capture the dominant coherent structures of the flow (figure~\ref{fig:fourier_schematic}).

\bibliographystyle{jfm}
\bibliography{jfm}
\end{document}